\documentclass[aps,preprint]{revtex4}%
\usepackage{amsfonts}
\usepackage{amsmath}
\usepackage{amssymb}
\usepackage{graphicx}
\usepackage{bm}%
\providecommand{\U}[1]{\protect\rule{.1in}{.1in}}

\begin{document}
\preprint{ }
\title[Guided Modes in the Plane Array]{Guided Modes in the Plane Array of Optical Waveguides}
\author{I. Ya. Polishchuk}
\affiliation{National Research Center Kurchatov Institute, Moscow, 123182, Russia\\
Max Planck Institute for the Physics of Complex Systems, D-01187 Dresden, Germany\\
Moscow Institute of Physics and Technology, Dolgoprudnii, 141700 Russia}
\author{A.A. Anastasiev}
\affiliation{National Research Nuclear University MEPhI
(Moscow Engineering Physics Institute), 115409, Kashirskoe shosse,
31, Moscow, Russia}
\author{M.I. Gozman}
\affiliation{Moscow Institute of Physics and Technology, Dolgoprudnii, 141700 Russia}
\author{Yu.I. Polishchuk}
\affiliation{National Research Center Kurchatov Institute, Moscow, 123182, Russia}
\author{S.V. Solov'ov}
\affiliation{Moscow Institute of Physics and Technology, Dolgoprudnii, 141700 Russia}
\author{E.A. Tsyvkunova}
\affiliation{National Research Nuclear University MEPhI
(Moscow Engineering Physics Institute), 115409, Kashirskoe shosse,
31, Moscow, Russia}

\keywords{Optical waveguides, guided modes, waveguide array, }
\pacs{42.81.Qb, 42.25.Bs, 42.82.Et, 63.20.Pw}

\begin{abstract}
It is known that, for the isolated dielectric cylinder waveguide, there
exists the cutoff frequency $\omega _{\ast }$ below which there are no
guided (radiationless) modes. It is shown in the paper that the infinite
plane periodic array of such waveguides possesses guided modes in the
frequency domain which is below the frequency $\omega _{\ast }$. So far as
the finite array is concerned, the modes in this frequency domain are weakly
radiating ones, but their quality factor $Q$ increases as $Q(N)\sim N^{3},$ $%
N$ ~being the number of the waveguides in the array. This dependence is
obtained both numerically, using the multiple scattering formalism, and is
justified within the framework a simple analytical model.
\end{abstract}
\maketitle

\bigskip


\section{Introduction}

\label{Sec_Intro}

The optical waveguides are the inherent component of the optical and
optoelectronic devices, indispensable for the optical signals transmission
between different parts of the system. The interaction between the closely
spaced waveguides usually results in the undesirable effects that distorts the
transmitted signal. However, in some cases, the interaction between the
waveguides can be exploited for practical purpose. In particular, this
concerns the plane periodic arrays of waveguides, which is a special kind of
low-dimensional photonic crystals. The main feature of such systems is a band
structure of the optical spectrum which defines its peculiar properties \cite%
{Lourtioz,Joannopoulos,Busch,Longhi}. Low-dimensional photonic crystal
composed of the parallel rods are of a special interest.
For the first time, the band structure of the plane array composed of the
semiconductor cylinders was investigated in Ref. \cite{maradudin1}. Then a
similar metallic structure which took into account dissipation was computed
in Ref. \cite{maradudin2}. Superconducting photonic crystals of such kind
were considered in \cite{lozovik}. Photonic crystals of such kind are useful
for various applications \cite{vergeles1}. In particular, they reveal
negative-angle refraction and reflection \cite{vergeles2}. On the other
hand, the arrays of parallel interacting cylinders is an ideal system to
simulate various physical phenomena inherent in condensed matter physics
such as Anderson localization, Bloch oscillations, Bloch-Zener tunneling,
etc. \cite{cite1, cite2, cite3, cite4, cite5}.

The electromagnetic filed which describes a guided mode in a waveguide is
finite inside the waveguide, while it is vanishing at a large distance from
it. It is well known that, for the isolated cylinder waveguide, the guided
modes can exist only above the so-called cutoff frequency $\omega _{\ast }$
\cite{Marcuse}. This is due to the fact that the conversion of the modes
with the frequency below $\omega _{\ast }$ into a free photon is possible.
This frequency depends on the material refractive index and the waveguide
diameter. However, in various applied tasks, it may be necessary to have a
guided mode below the frequency $\omega _{\ast }$ for the given waveguides.
In particular, such problem may appear in the following case. Indeed, along
with the radiation, there exist losses connected with the absorption by the
material itself. If the frequency window is located below the cutoff
frequency $\omega _{\ast }$, there arise a problem to shift the frequency $%
\omega _{\ast }$ into this window region.

In this paper, we investigate a possibility of a formation of high quality
guided modes in the plane waveguide array which are below the cutoff
frequency of the isolated waveguide $\omega _{\ast }$. Thus our aim is to
suppress the radiation loss using the array of the waveguides. First, we
consider the \textit{infinite} plane periodical array of the waveguides an
show that, taking into account the interaction between them, results in the
appearance of the guided modes below $\omega _{\ast }$. These modes possess
the infinite $Q-$factor. However, actually we deal with the array of a
finite size and the modes become low-radiating ones. For this reason, we
investigate how the $Q-$factor of the low-radiating modes depends on the
number of the waveguides in the array $N$. Using the multiple scattering
formalism (MSF), it is shown numerically that $Q(N)\sim N^{3}$. This
formalism is based on the exact description of the electromagnetic waves which are scattered by the infinite cylinder \cite{VanDeHulst}. Besides, we propose a
simple model which qualitatively explains this cubic dependence.

Note that the effect of increasing the $Q-$factor for the radiative modes
in the array of the interacting spherical particles with increasing the array size, was
discovered in \cite{We-OptExpr2007, Burin-PhysRevE2006, deich,wePRE2010}.

The paper is organized as follows. In Section \ref{Sec_MSF} we describe the
MSF giving a brief derivation of the principal relations.
Based on the relations obtained, the numerical simulation for the infinite and the finite arrays of cylinder waveguides is given in  Section \ref{Sec_NumSim}. A clear qualitative explanation of the results obtained numerically is given in Section \ref{theory}.


\section{Multiple scattering formalism}

\label{Sec_MSF}

\begin{figure}[ptbh]
\centering
\includegraphics[width=0.5\textwidth]{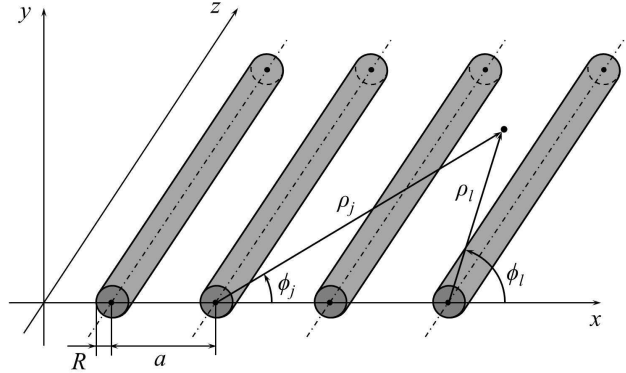}
\caption{The optical waveguide array. The polar coordinates of radius-vector
$\mathbf{r}$ relative to different waveguides.}
\label{fig1a}
\end{figure}

Let us consider the array of $N$ parallel cylindrical dielectric waveguides
(see figure \ref{fig1a}). The axes of the waveguides are in the $xz$-plane
and are parallel to the $z$-axis. The array is equidistant, $a$ being the
distance between the axes of the nearest waveguides. All the waveguides are
assumed to have the same radius $R$ and the same refractive index $n$. The
refractive index of the environment is $n_{0}$. The system of units where
the speed of light $c=1$ is used.

Let a guided mode with a frequency $\omega $ is excited. Because of the
translation invariance in the $z$ direction, all the components of the
electromagnetic field describing the guided mode depend on the coordinate $z$
as $e^{i\beta z}$, $\beta $ being a propagation constant. Thus, all the
components of the electromagnetic field describing the guided mode are
proportional to the factor $e^{-i\omega t+i\beta z}$. First, let us describe
the electromagnetic field inside the waveguides. The field inside the $j$-th
waveguide, being of a finite value, may be represented as follows
\begin{equation}
\begin{array}{l}
\tilde{\mathbf{E}}_{j}(t,\mathbf{r})=e^{-i\omega t+i\beta
z}\,\sum\limits_{m=0,\pm 1...}e^{im\phi _{j}}\,\Bigl(c_{jm}\,\tilde{\mathbf{N%
}}_{\omega ^{\prime }\beta m}(\rho _{j})-d_{jm}\,\tilde{\mathbf{M}}_{\omega
^{\prime }\beta m}(\rho _{j})\Bigr),\medskip \\
\tilde{\mathbf{H}}_{j}(t,\mathbf{r})=e^{-i\omega t+i\beta
z}\,n\sum\limits_{m=0,\pm 1...}e^{im\phi _{j}}\,\Bigl(c_{jm}\,\tilde{\mathbf{%
M}}_{\omega ^{\prime }\beta m}(\rho _{j})+d_{jm}\,\tilde{\mathbf{N}}_{\omega
^{\prime }\beta m}(\rho _{j})\Bigr),%
\end{array}
\label{AppS_EHint}
\end{equation}%
where $\mathbf{r}=(x,y,z)=(\bm{\rho},z)$, $\rho _{j}=|\bm{\rho}-\mathbf{a}j|<R$, $%
\phi _{j}$ is the polar angle for the vector $\bm{\rho}-\mathbf{a}j$ (see figure %
\ref{fig1a}), $\omega ^{\prime }=n\omega $. The vector cylinder harmonics $%
\tilde{\mathbf{M}}_{\omega ^{\prime }\beta m}(\rho _{j})$ and $\tilde{%
\mathbf{N}}_{\omega ^{\prime }\beta m}(\rho _{j})$ are defined as follows
\begin{equation}
\tilde{\mathbf{N}}_{\omega ^{\prime }\beta m}(\rho _{j})=\mathbf{e}_{r}\,%
\frac{i\beta }{\kappa }\,J_{m}^{\prime }(\kappa \rho _{j})-\mathbf{e}_{\phi
}\,\frac{m\beta }{\kappa ^{2}\rho _{j}}\,J_{m}(\kappa \rho _{j})+\mathbf{e}%
_{z}\,J_{m}(\kappa \rho _{j}),  \label{tildeN}
\end{equation}%
\begin{equation}
\tilde{\mathbf{M}}_{\omega ^{\prime }\beta m}(\rho _{j})=\mathbf{e}_{r}\,%
\frac{m\omega ^{\prime }}{\kappa ^{2}\rho _{j}}\,J_{m}(\kappa \rho _{j})+%
\mathbf{e}_{\phi }\,\frac{i\omega ^{\prime }}{\kappa }\,J_{m}^{\prime
}(\kappa \rho _{j}),  \label{tildeM}
\end{equation}%
here $\kappa =\sqrt{n^{2}\omega ^{2}-\beta ^{2}}$, $J_{m}(\kappa \rho _{j})$
is the Bessel function, and the prime means the derivative with respect to
the argument $\kappa \rho _{j}$. Thus, the guided mode inside the $j$-th
waveguide is determined by the frequency $\omega $, by the propagation
constant $\beta $ and by the partial amplitudes $c_{jm}$, $d_{jm}$.

Now let us turn to the electromagnetic field for the same guided mode
outside of the array. This field is a sum of the contributions induced by
all the waveguides:
\begin{equation}
\mathbf{E}(t,\mathbf{r})=\sum\limits_{j=1}^N\mathbf{E}_j(t,\mathbf{r}),
\qquad \mathbf{H}(t,\mathbf{r})=\sum\limits_{j=1}^N\mathbf{H}_j(t,\mathbf{r}%
).  \label{Sum11}
\end{equation}
For the guided mode, the contribution of the $j$-th waveguide should vanish
at $\rho_j\to\infty$. Therefore, one can write
\begin{equation}
\begin{array}{l}
\mathbf{E}_j(t,\mathbf{r})=e^{-i\omega t+i\beta z}\sum\limits_{m}e^{i
m\phi_j}\,\Bigl(a_{jm}\,\mathbf{N}_{\omega_0\beta m}(\rho_j)-b_{jm}\,\mathbf{%
M}_{\omega_0\beta m}(\rho_j )\Bigr),\medskip \\
\mathbf{H}_j(t,\mathbf{r})=e^{-i\omega t+i\beta z}n_0 \sum\limits_me^{i
m\phi_j}\,\Bigl(a_{jm}\,\mathbf{M}_{\omega_0\beta m}(\rho_j)+b_{jm}\,\mathbf{%
N}_{\omega_0\beta m}(\rho_j)\Bigr).%
\end{array}
\label{EX_J}
\end{equation}
where $\rho_j>R$. In (\ref{EX_J}), another kind of the vector cylinder
harmonics is introduced
\begin{equation}
\mathbf{N}_{\omega_0\beta m}(\rho_j)=\mathbf{e}_r\, \frac{i\beta}{\kappa_0}%
\,H^{\prime }_m(\kappa_0\rho_j) -\mathbf{e}_\phi\,\frac{m\beta}{%
\kappa_0^2\rho_j}\,H_m (\kappa_0\rho_j)+\mathbf{e}_z\,H_m(\kappa_0\rho_j),
\label{N}
\end{equation}
\begin{equation}
\mathbf{M}_{\omega_0\beta m}(\rho_j)=\mathbf{e}_r\,\frac {m\omega_0}{%
\kappa_0^2\rho_j}\,H_m(\kappa_0\rho_j)+\mathbf{e}_\phi\,\frac{i\omega_0}{%
\kappa_0}\,H^{\prime }_m(\kappa_0\rho_j),  \label{M}
\end{equation}
where $H_{m}(\kappa_0\rho_j)$ is the Hankel function of the first kind, $%
\omega_0=n_0\omega$, $\kappa_0=\sqrt{n_0^2\omega^2-\beta^2}$. Thus, the
contribution of the $j$-th waveguide into the guided mode field outside of
the array is determined by the frequency $\omega$, by the propagation
constant $\beta$, by the partial amplitudes $a_{jm}$, $b_{jm}$. 
Note that, for $\beta=0$, expansions (\ref{AppS_EHint}) and (\ref{EX_J})
transform into the corresponding expressions in \cite{VanDeHulst}, however
different notations are used there. Below, the factor $e^{-i\omega t+i\beta
z}$ is omitted.

Below in this paper, for the purpose of illustration of the effect, we
confine ourselves to the zero-harmonic approximation. This means that only
the terms with $m=0$ are taken into account in (\ref{AppS_EHint}) and (\ref%
{EX_J}). It is easy to convince ourselves that in this case the guided modes
are either transverse magnetic modes (TM) or transverse electric (TE) ones.
For the TM mode $b_{j0}=d_{j0}=0$, while for the TE mode $a_{j0}=c_{j0}=0$.
As an example, let us consider the TM modes. Then, equations (\ref%
{AppS_EHint}) and (\ref{EX_J}) take the form
\begin{equation}
\tilde{\mathbf{E}}_{j}(\mathbf{r})=c_{j}\,\tilde{\mathbf{N}}_{\omega
^{\prime }\beta 0}(\rho _{j}),\qquad \tilde{\mathbf{H}}_{j}(\mathbf{r}%
)=c_{j}n\,\tilde{\mathbf{M}}_{\omega ^{\prime }\beta 0}(\rho _{j}),\qquad
\rho _{j}<R  \label{MSF_101a}
\end{equation}%
\begin{equation}
\mathbf{E}_{j}(\mathbf{r})=a_{j}\,\mathbf{N}_{\omega _{0}\beta 0}(\rho
_{j}),\qquad \mathbf{H}_{j}(\mathbf{r})=a_{j}n_{0}\,\mathbf{M}_{\omega
_{0}\beta 0}(\rho _{j}),\qquad \rho _{j}>R  \label{MSF_101b}
\end{equation}%
Here the notations $a_{j}$, $c_{j}$ are used instead of $a_{j0}$, $c_{j0}$.

Let $\mathbf{R}_{j}$ be the radius-vector of a point on the surface of the $%
j $-th waveguide. Then the fields $\tilde{\mathbf{E}}_{j}(\mathbf{R}_{j})$, $%
\tilde{\mathbf{H}}_{j}(\mathbf{R}_{j})$ in Eq.(\ref{AppS_EHint}) and the
fields $\mathbf{E}(\mathbf{R}_{j})$ $\mathbf{H}(\mathbf{R}_{j})$ in Eq.(\ref%
{Sum11}) obey the boundary conditions on the surface of this waveguide.
Generally, there are six boundary conditions. However, for the TM-modes only
two of them are required.
\begin{equation}
\lbrack \mathbf{E}(\mathbf{R}_{j})]_{z}=[\tilde{\mathbf{E}_{j}}(\mathbf{R}%
_{j})]_{z},\qquad \lbrack \mathbf{H}(\mathbf{R}_{j})]_{\phi }=[\tilde{%
\mathbf{H}_{j}}(\mathbf{R}_{j})]_{\phi }.  \label{MSF_BoundCond}
\end{equation}%
These equations determine the partial amplitude $a_{j}$ and $c_{j}$ for
given $\omega $ and $\beta .$ To represent Eqs.(\ref{MSF_BoundCond}) in a
convenient form, one should express the fields $\mathbf{E}_{l}(\mathbf{R}%
_{j})$ for $l\neq j$ entering in Eq.(\ref{Sum11}) in terms of the functions $%
\tilde{\mathbf{N}}$ using the Graph formula
\begin{equation}
\left[
\begin{array}{c}
\mathbf{N}_{\omega _{0}\beta n}(\rho _{l}) \\
\mathbf{M}_{\omega _{0}\beta n}(\rho _{l})%
\end{array}%
\right] \,e^{in\phi _{l}}=\sum\limits_{m=0}^{+\infty }U_{nm}^{lj}(\omega
,\beta )\,\left[
\begin{array}{c}
\tilde{\mathbf{N}}_{\omega _{0}\beta m}(\rho _{j}) \\
\tilde{\mathbf{M}}_{\omega _{0}\beta m}(\rho _{j})%
\end{array}%
\right] \,e^{im\phi _{j}},
\end{equation}%
here
\begin{equation}
U_{nm}^{lj}(\omega ,\beta )=H_{n-m}(\kappa _{0}a|l-j|)\,[\mathrm{sign}%
(j-l)]^{n-m}.  \label{MSF_U}
\end{equation}%
For the $m=0$ approximation one has
\begin{equation}
\begin{array}{l}
\mathbf{N}_{\omega _{0}\beta 0}(\rho _{l})\approx U_{l-j}(\omega ,\beta )\,%
\tilde{\mathbf{N}}_{\omega _{0}\beta 0}(\rho _{j}), \\
\mathbf{M}_{\omega _{0}\beta 0}(\rho _{l})\approx U_{l-j}(\omega ,\beta )\,%
\tilde{\mathbf{M}}_{\omega _{0}\beta 0}(\rho _{j}),%
\end{array}
\label{MSF_110}
\end{equation}%
where $U_{l-j}(\omega ,\beta )=U_{00}^{lj}(\omega ,\beta )$. Then, it
follows from Eq.(\ref{MSF_101b}) that
\begin{equation}
\begin{array}{l}
\mathbf{E}_{l}(\mathbf{r})=a_{l}\,U_{l-j}(\omega ,\beta )\,\tilde{\mathbf{N}}%
_{\omega _{0}\beta 0}(\rho _{j}), \\
\mathbf{H}_{l}(\mathbf{r})=a_{l}\,n_{0}\,U_{l-j}(\omega ,\beta )\,\tilde{%
\mathbf{M}}_{\omega _{0}\beta 0}(\rho _{j}).%
\end{array}
\label{MSF_115}
\end{equation}%
Thus,
\begin{equation}
\begin{array}{l}
\mathbf{E}(\mathbf{r})=a_{j}\,\mathbf{N}_{\omega _{0}\beta 0}(\rho
_{j})+\sum\limits_{l\neq j}a_{l}\,U_{l-j}(\omega ,\beta )\,\tilde{\mathbf{N}}%
_{\omega _{0}\beta 0}(\rho _{j}), \\
\mathbf{H}(\mathbf{r})=a_{j}\,n_{0}\,\mathbf{M}_{\omega _{0}\beta 0}(\rho
_{j})+\sum\limits_{l\neq j}a_{l}\,n_{0}\,U_{l-j}(\omega ,\beta )\,\tilde{%
\mathbf{M}}_{\omega _{0}\beta 0}(\rho _{j}).%
\end{array}
\label{MSF_120}
\end{equation}%
Substituting (\ref{MSF_120}) and (\ref{MSF_101a}) into (\ref{MSF_BoundCond}%
), one obtains
\begin{equation}
\begin{array}{l}
a_{j}\,H_{0}(\kappa _{0}R)+\sum\limits_{l\neq j}a_{l}\,U_{l-j}(\omega ,\beta
)\,J_{0}(\kappa _{0}R)=c_{j}\,J_{0}(\kappa R), \\
a_{j}\,n_{0}\,\frac{i\omega _{0}}{\kappa _{0}}H_{0}^{\prime }(\kappa
_{0}R)+\sum\limits_{l\neq j}a_{l}\,n_{0}\,U_{l-j}(\omega ,\beta )\frac{%
i\omega _{0}}{\kappa _{0}}J_{0}^{\prime }(\kappa _{0}R)=c_{j}\,n\,\frac{%
i\omega ^{\prime }}{\kappa }J_{0}^{\prime }(\kappa R).%
\end{array}
\label{MSF_130}
\end{equation}%
Then the system of equations (\ref{MSF_130}) is reduced to the form
\begin{equation}
\frac{a_{j}}{\bar{a}(\omega ,\beta )}-\sum\limits_{l\neq j}U_{l-j}(\omega
,\beta )\,a_{l}=0,  \label{MSF_Main}
\end{equation}%
\begin{equation}
c_{j}=\bar{c}(\omega ,\beta )\,a_{j},  \label{MSF_140}
\end{equation}%
where
\begin{equation}
\bar{a}(\omega ,\beta )=\frac{n^{2}\kappa _{0}\,J_{0}^{\prime }(\kappa
R)\,J_{0}(\kappa _{0}R)-n_{0}^{2}\kappa \,J_{0}(\kappa R)\,J_{0}^{\prime
}(\kappa _{0}R)}{n_{0}^{2}\kappa \,J_{0}(\kappa R)\,H_{0}^{\prime }(\kappa
_{0}R)-n^{2}\kappa _{0}\,J_{0}^{\prime }(\kappa R)\,H_{0}(\kappa _{0}R)},
\label{MSF_a}
\end{equation}%
\begin{equation}
\bar{c}(\omega ,\beta )=\frac{n_{0}^{2}\kappa \{H_{0}(\kappa
_{0}R)J_{0}^{\prime }(\kappa _{0}R)-H_{0}^{\prime }(\kappa
_{0}R)J_{0}(\kappa _{0}R)\}}{n_{0}^{2}\kappa J_{0}^{\prime }(\kappa
_{0}R)J_{0}(\kappa R)-n^{2}\kappa J_{0}(\kappa _{0}R)J_{0}^{\prime }(\kappa
R)}.  \label{MSF_c}
\end{equation}%
The terms $U_{l-j}(\omega ,\beta )$ in Eq.(\ref{MSF_Main}) desribe the
interaction between the waveguides. If the terms $U_{l-j}(\omega ,\beta )$
in Eq.(\ref{MSF_Main}) are neglected, the poles of $\bar{a}(\omega ,\beta )$
or $\bar{c}(\omega ,\beta )$ determines the guied modes for the isolated
wavegide.


System of equations (\ref{MSF_Main}) possesses nontrivial solutions if
\begin{equation}
\det\left\|\frac{\delta_{jl}}{\bar{a}(\omega,\beta)}-U_{l-j}(\omega,\beta)%
\right\|=0.  \label{Eq005}
\end{equation}
This equation relates the frequency of the guided mode $\omega$ and its
propagation constant $\beta$ implicitly.

For the infinite periodical array of identical waveguides, the solution of
Eq.(\ref{MSF_Main}) reads%
\begin{equation}
a_{j}=a_{0}\,e^{ikaj},\qquad -\pi /a<k\leq \pi /a.  \label{Eq006}
\end{equation}%
In this case, the nontrivial solution exists if
\begin{equation}
\frac{1}{\bar{a}(\omega ,\beta )}-U(\omega ,\beta ,k)=0,  \label{Eq007}
\end{equation}%
where
\begin{equation}
U(\omega ,\beta ,k)=\sum\limits_{l\neq 0}U_{l}(\omega ,\beta )\,e^{ikal}.
\label{Eq008}
\end{equation}%
Equation (\ref{Eq007}) determines the dispersion law $\omega (\beta ,k)$. If
the propagation constant $\beta $ is real, the corresponding mode
frequencies may be either real or complex. If the frequency is real, the
mode possesses an infinite $Q-$factor. Otherwise the mode has a finite
lifetime and the imaginary part of the frequency determines the mode decay
rate. However, if the corresponding quality factor is large, the mode may be
considered as a guided one.

\section{Numerical simulation for the infinite and the finite arrays.}

\label{Sec_NumSim}

Let us consider the infinite the array of the waveguides with the geometric
parameters and the refractive indices which are chosen to be close to those
in Refs. \cite{cite5}. The specific values of the parameters are taken so
that the illustration of the results looks quite representative. For this
reason, one takes the waveguide radius $R=1.975\mathrm{\mu m}$, the
refractive index of the waveguides $n=1.554$, and the refractive index of
the environment $n_{0}=0.99n=1.538$. Using (\ref{MSF_a}) one finds the
cutoff frequency for the isolated waveguide $\omega _{\ast }=5.57\mathrm{\mu
m}^{-1}.$ This value corresponds to the cutoff propagation constant $\beta
_{\ast }=8.57\mathrm{\mu m}^{-1}$. (Let us remind that the speed of light $%
c=1$ and, therefore, the frequency has a dimension of the inverse length).

\begin{figure}[tbph]
\centering
\includegraphics[width=0.5\textwidth]{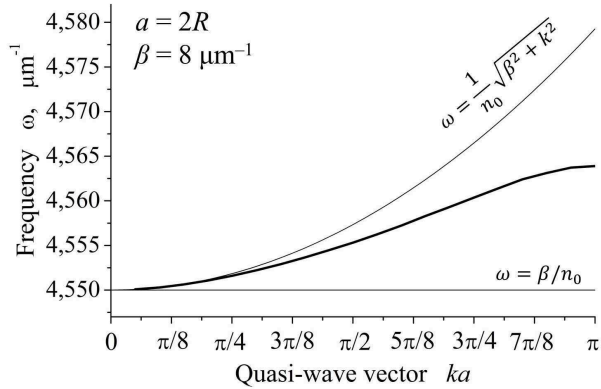} 
\caption{The dispersion curve for the infinite array.}
\label{Fig2}
\end{figure}

The modes we are interested in, appear due to the interaction between the
waveguides. For this reason, we assume that the neighbor waveguides touch
each other, since in this case the interaction reveals itself the most
strongly. As an example, let us choose the propagation constant $\beta =8%
\mathrm{\mu m}^{-1}<\beta _{\ast }$. The dispersion curve $\omega (\beta
,k), $which is a numerical solution to Eq. (\ref{Eq007}), is presented in
Fig. \ref{Fig2} by the thick solid line.

One sees that the curve is completely located within the domain $\beta
<n_{0}\omega (k)<\sqrt{\beta ^{2}+k^{2}}.$ A physical explanation to this
fact is given below. This numerical results completely supports the
kinematic criterion for the mode to be a radiationless one. Thus the
infinite periodical array of the waveguides may possess the guides modes
with the frequencies below the cutoff frequency of the single waveguide.

The radiationless guided modes inherent in the periodical array found above
(see Fig.\ref{Fig2}) possess the \textit{infinite }quality factor $Q=2%
\mathrm{Re}\,\omega /|\mathrm{Im}\,\omega |$. This is due to the fact that
the array is infinite. However, actually one deals with the arrays composed
of a finite number of the waveguides $N$. On the other hand, it is evident
that for $N\gg 1$ the array should manifest the features similar to the
infinite array. In particular, the guided modes should possess a high
quality. Let us investigate how the quality factor $Q$ depends on the number
of the waveguides $N.$

\begin{figure}[tbp]
\includegraphics[width=0.4\textwidth]{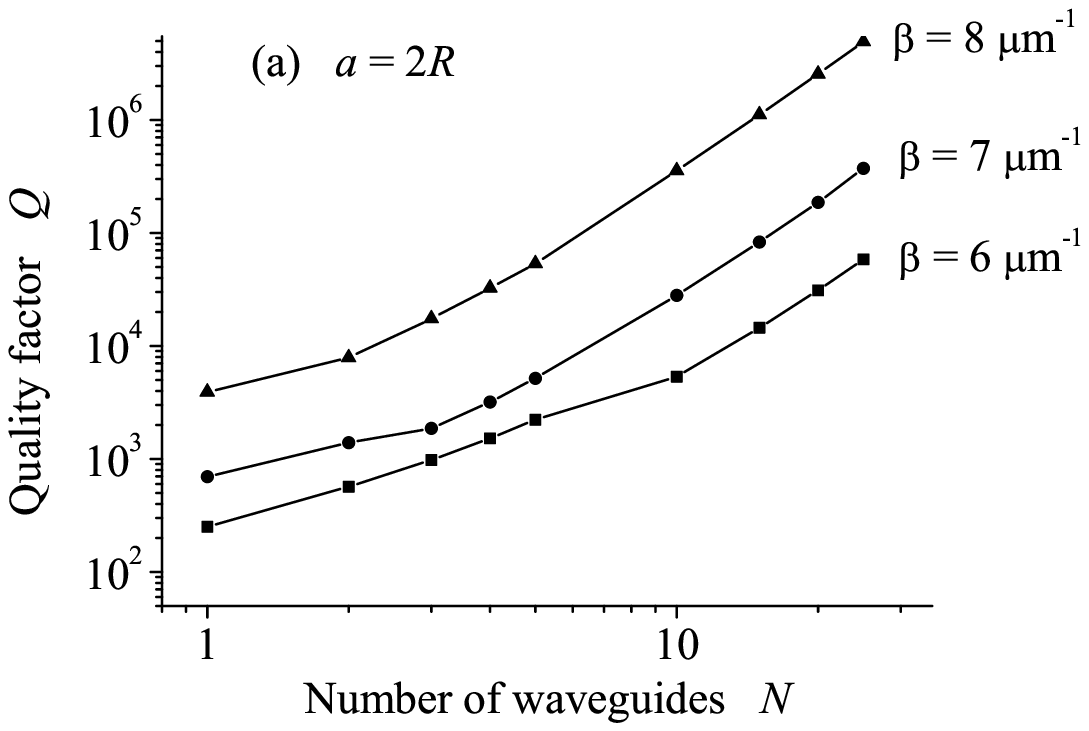} \hfill %
\includegraphics[width=0.4\textwidth]{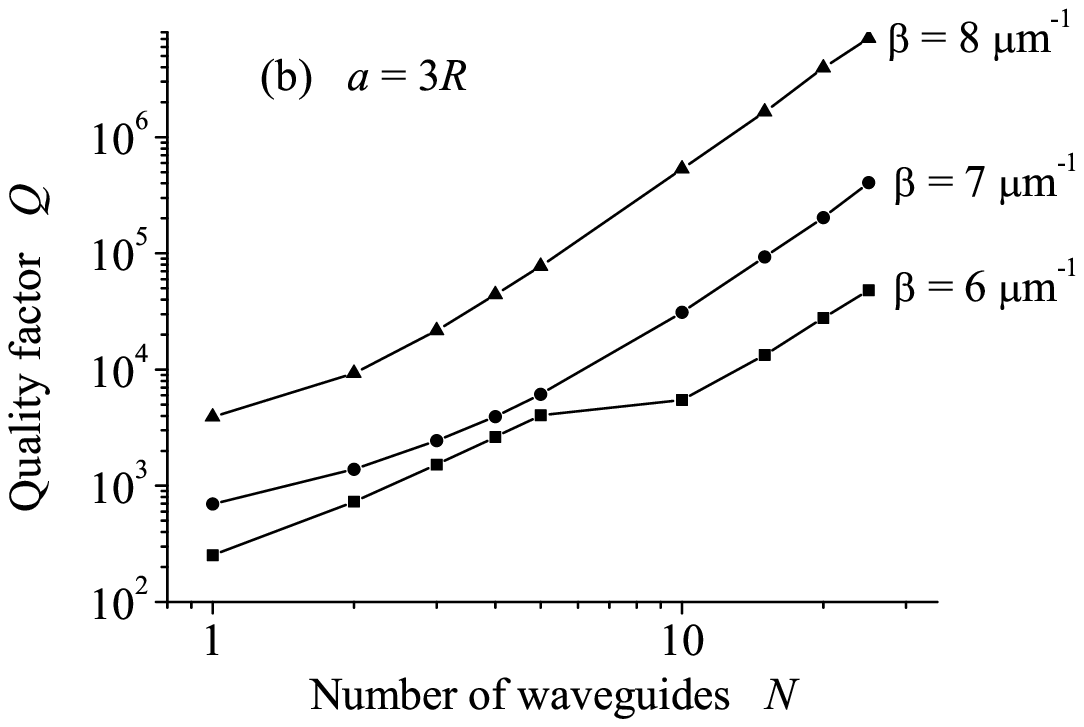}
\caption{The dependence of the quality factor on the number of the
waveguides $N$ for the plane array of waveguides.}
\label{Fig4}
\end{figure}

Using Eq. (\ref{Eq005}) one can obtain numerically that, for a finite $N\gg
1,$ the highest quality factor is reached for the modes whose frequency is
close to the upper edge of the Brillouin zone $k\approx \pi /a$ (such a
feature is inherent also for the array of spherical particles \cite%
{We-OptExpr2007, Burin-PhysRevE2006,wePRE2010}). The dependence of the
quality factor on the number of the waveguides $N$ just for modes with the
highest $Q-$factor is illustrated by two example: the waveguides are
touching, $a=2R,$ and the waveguides are spatially separated, $a=3R$. The
three values for the propagation constant smaller than $\beta _{\ast }$ are
taken: $\beta _{1}=6\mathrm{\mu m}^{-1}$, $\beta _{2}=7\mathrm{\mu m}^{-1}$,
$\beta _{3}=8\mathrm{\mu m}^{-1}<\beta _{\ast }$. The results of the
numerical simulation are presented in Fig. \ref{Fig4}. The analysis of the
dependencies in this figure reveals a remarkable feature: for $N>10$ the
dependence the quality factor $Q(N)\sim N^{3}$.

\begin{figure}[tbph]
\centering
\includegraphics[width=0.5\textwidth]{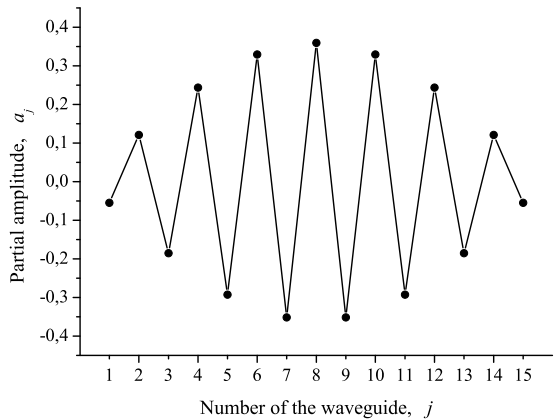}
\caption{The partial amplitude for the mode with the highest quality factor.}
\label{Fig3a}
\end{figure}

Let us retrieve, using Eq. (MSF\_Main), the relation between the partial
amplitudes $a_{j}.$ A typical dependence, obtained numerically, is
presented, as an example, in Fig. \ref{Fig3a} for the case $N=15.$

\section{The interpretation of the numerical results}
\label{theory}

Knowing the dependence $\omega (\beta ,k)$ allows us to determine the
features of the guided modes. Let us pay attention, that the reason for the
mode to possesses a finite lifetime is a conversion of it into a free
photon. That is the mode is a radiative one. First, let us consider a single
waveguide. Then the mode is described by a frequency $\omega $ and a
propagation constant $\beta $. For the conversion into a free photon to
takes place, the photon wave vector $\mathbf{q}$ should satisfy the two
conditions: $|\mathbf{q}|=n_{0}\omega $ and $q_{z}=\beta $. Since $q_{z}<|%
\mathbf{q}|$, the photon can be emitted \textit{only} if $\beta <n_{0}\omega
$. In the opposite case $\beta >n_{0}\omega $, the mode is a radiationless
one and it is a guided one with the infinite quality factor. Let us turn to
the infinite plane periodic array. In this case, a mode is determined by a
quasi-wave vector $k$, in addition to the frequency $\omega $ and the
propagation constant $\beta $. Thus, the wave vector $\mathbf{q}$ of the
emitted free photon satisfies three conditions: $|\mathbf{q}|=n_{0}\omega $,
$q_{z}=\beta $ and $q_{x}=k$. It is obvious that $\sqrt{q_{x}^{2}+q_{z}^{2}}%
<|\mathbf{q}|$. So, the mode can be converted into a free photon only if $%
\sqrt{k^{2}+\beta ^{2}}<n_{0}\omega $. Thus, the infinite periodical array
possesses guided modes within the frequency domain, which obeys the
kinematic criterion
\begin{equation}
\beta <n_{0}\omega <\sqrt{\beta ^{2}+k^{2}},  \label{citerion}
\end{equation}%
where the single waveguide allow only the radiating modes. Note that, since $%
n\omega (\beta ,k)<\sqrt{\beta ^{2}+k^{2}},$ a guided mode may not exist for
small $k$ at all.

Then, let us explain qualitatively the cubic dependence for the $Q$-factor
found in the previous section. First, let us consider the infinite array of
the waveguides. Let $A_{j}(t)$ be the effective time-dependent partial
amplitude for the $j$-th waveguide, which characterizes the waveguide as a
whole. The time evolution of $A_{j}(t)$ may be approximately described by
the equation which similar to a Schr\"{o}dinger one
\begin{equation}
i\frac{dA_{j}}{dt}(t)+V\Bigl(A_{j-1}(t)+A_{j+1}(t)\Bigr)=0.  \label{Der010}
\end{equation}%
Here $V$ is the effective coupling between the nearest waveguides. Let us
find the solution for this equation in the form
\begin{equation}
A_{j}(t)=A_{0}\,e^{-i\omega t+ikj},\qquad -\pi <k\leq \pi .  \label{Der020}
\end{equation}%
Substituting (\ref{Der020}) into (\ref{Der010}) one obtains the dispersion
law:
\begin{equation}
\omega (k)=-2V\cos k.  \label{Der030}
\end{equation}%
\qquad

Now let us turn to the finite array. As found above, the infinite array
possesses the infinite $Q,$ while the finite array possesses a large but a
finite $Q.$ (see Fig. \ref{Fig4}). For this reason, it is natural to assume
that this is connected with the availability of the \textit{edge }waveguides
in the array which are responsible for the radiation of the photon. Based on
this fact, one can write for the finite array the equation similar to Eq. (%
\ref{Der010}):%
\begin{equation}
i\frac{dA_{j}}{dt}(t)+V\,\Bigl(1-\delta _{j1}\Bigr)\,A_{j-1}(t)+V\,\Bigl(%
1-\delta _{jN}\Bigr)\,A_{j+1}(t)+i\gamma \,\Bigl(\delta _{j1}+\delta _{jN}%
\Bigr)\,A_{j}(t)=0.  \label{Der040}
\end{equation}%
The parameter $\gamma \ll V$ is responsible for the free photon emission.
Using (\ref{Der020}) one obtains from (\ref{Der040})
\begin{equation}
\omega \,A_{j}+V(1-\delta _{j1})\,A_{j-1}+V(1-\delta _{jN})\,A_{j+1}+i\gamma
(\delta _{j1}+\delta _{jN})\,A_{j}=0.  \label{Der050}
\end{equation}%
Note that $\omega =\omega ^{\prime }+i\gamma $ may be complex. For the
particular $j=N,$ this equation takes the form
\begin{equation}
(\omega ^{\prime }+i\gamma )\,A_{N}+V\,A_{N-1}=0.  \label{Der060}
\end{equation}%
So,
\begin{equation}
\omega ^{\prime }+i\gamma =-V\,\frac{A_{N-1}}{A_{N}}.  \label{Der070}
\end{equation}

The dependence for $a_{j}$ in Fig. \ref{Fig3a} approaches zero at the edges
of the array and resembles a cosine one. For this reason, let us seek the
solution to Eq. (\ref{Der050}) in the form
\begin{equation}
A_{j}\sim \cos k\,(j-N/2),  \label{Der080}
\end{equation}%
$k$ being close to $\pi -\pi /N$. Substituting (\ref{Der030}) and (\ref%
{Der080}) into (\ref{Der070}), one gets
\begin{equation}
-2V\,\cos \,k+i\gamma =-V\frac{\cos \,k(N/2-1)}{\cos \,kN/2}.  \label{Der090}
\end{equation}%
Let
\begin{equation}
k=\pi -\pi /N+x,  \label{Der095}
\end{equation}%
where $x$ is complex and $|x|\ll \pi /N$. For the sake of simplicity, let us
assume $N$ to be even. Then, substituting (\ref{Der095}) into (\ref{Der090}%
), one gets
\begin{equation}
2V\,\cos \,\left( \frac{\pi }{N}-x\right) +i\gamma =V\,\frac{\sin \,(\pi
/N+Nx/2)}{\sin \,Nx/2}.  \label{Der100}
\end{equation}%
Taking into account a smallness of the arguments in the trigonometric
functions in (\ref{Der100}) one obtains:
\begin{equation}
2V+i\gamma \approx V\,\left( \frac{2\pi }{N^{2}x}+1\right) .  \label{Der110}
\end{equation}%
Then, since $\gamma \ll V$, one has
\begin{equation}
x\approx \frac{2\pi }{N^{2}}\,\left( 1-i\frac{\gamma }{V}\right) .
\label{Der120}
\end{equation}%
Substituting (\ref{Der120}) and (\ref{Der095}) into (\ref{Der030}), one
gets:
\begin{equation}
\omega \approx 2V-i\frac{4\pi ^{2}\gamma }{N^{3}}.  \label{Der130}
\end{equation}%
Then, the quality factor
\begin{equation}
Q=\frac{2\mathrm{Re}\,\omega }{|\mathrm{Im}\,\omega |}=\frac{VN^{3}}{\pi
^{2}\gamma }.  \label{Der140}
\end{equation}%
reveal the sought-for cubic dependence.

\section{Conclusion}

\label{Sec_Conclusion}

In this paper we investigated the guided modes in the array of coupled
waveguides below the cutoff frequency of a single waveguide, i.e. in the
frequency domain where the single waveguide possesses only the radiating
modes. It is shown that the infinite periodic array possesses a band of the
guided modes with the infinite $Q$-factor. In the case of the finite array,
the modes below the cutoff frequency are weakly radiating ones. Their
quality factor increases with the number of waveguides as $Q(N)\sim N^{3}.$%
These results are obtained numerically using the multiple scattering
formalism. A clear physical interpretation of the numerical results is given.

\section*{Acknowledgements}

\label{Sec_Acknowledgements}

The study is partially supported by the Russian Fund for Basic Research
(Grant 16-02-00660).


\end{document}